\documentclass[11pt,a4paper]{article}

\usepackage{graphicx}
\usepackage{bm}
\usepackage{amsmath} 
\usepackage{braket}
\usepackage{ragged2e} 
\usepackage{placeins} 

\usepackage{caption}
\usepackage{subcaption}
\usepackage[labelfont=bf]{caption}
\usepackage{hyperref}
\hypersetup{
  colorlinks=true,
  linkcolor=black,
  citecolor=blue,
  urlcolor=blue
}

\usepackage[
  left=2.5cm,
  right=2.5cm,
  top=2.5cm,
  bottom=2.5cm
]{geometry}

\usepackage[backend=biber,style=numeric,sorting=none]{biblatex}
\def\papertitle{
	Electrical Spin Pumping in Exchange-coupled Molecules
}

\newcommand{\AuthorList}{\normalsize
    Paul~Greule$^{1}$,
	Wantong~Huang$^{1}$,
	Kwan~Ho~Au-Yeung$^{1,2}$,
    Máté~Stark$^{1}$, \\ \normalsize
    Johannes~Schwenk$^{1}$,
    Christoph~Sürgers$^{1,3}$,
    Wolfgang~Wernsdorfer$^{1,2,3}$,
    Philip~Willke$^{1,2\ast}$\\
}

\newcommand{\Affiliations}{%
\small
$^{1}$Physikalisches Institut (PHI), Karlsruhe Institute of Technology (KIT), Karlsruhe, Germany.\\
$^{2}$Center for Integrated Quantum Science and Technology (IQST), Karlsruhe Institute of Technology, Karlsruhe, Germany.\\
$^{3}$Institute for Quantum Materials and Technologies (IQMT), Karlsruhe, Germany.\\
\normalsize
}

\newcommand{\CorrespondingAuthor}{%
\small $^{*}$corresponding author: philip.willke@kit.edu \normalsize
}

\usepackage{chngcntr}

\newcommand{\beginsupplement}{%
  \clearpage

  \setcounter{page}{1}      
  \setcounter{section}{0}
  \setcounter{subsection}{0}
  \setcounter{subsubsection}{0}
  \setcounter{figure}{0}
  \setcounter{table}{0}
  \setcounter{equation}{0}

  \renewcommand{\thepage}{S\arabic{page}}
  \renewcommand{\thesection}{\arabic{section}}
  \renewcommand{\thesubsection}{\arabic{section}.\arabic{subsection}}
  \renewcommand{\thesubsubsection}{\arabic{section}.\arabic{subsection}.\arabic{subsubsection}}
  \renewcommand{\thefigure}{S\arabic{figure}}
  \renewcommand{\thetable}{S\arabic{table}}
  \renewcommand{\theequation}{S\arabic{equation}}

  \counterwithout{equation}{section}
  \counterwithout{figure}{section}
  \counterwithout{table}{section}
}

\makeatletter
\newcommand{\maketitlesupplement}{%
  \clearpage
  \beginsupplement

  \let\old@title\@title
  \let\old@author\@author
  \let\old@date\@date

  \title{{\large Supplementary Information of\par}\vspace{0.6em}\PaperTitle}%
  \author{\AuthorList}%
  \date{\today}%

  \maketitle
  \Affiliations

  \let\@title\old@title
  \let\@author\old@author
  \let\@date\old@date
}
\makeatother

\title{\bfseries \boldmath \papertitle}
\author{\AuthorList}
\begin{document}
\date{}
\maketitle
\noindent \Affiliations \\
\CorrespondingAuthor

\vspace{2cm}

\begin{abstract}
\justifying
\noindent \textbf{Electron spins in single molecules are a promising platform for quantum information processing. However, their practical implementation as qubits requires reliable control at the single-entity level, including an efficient state initialization. Here, we demonstrate the remote, all-electrical initialization of the electron spin in single molecules: Using electron spin resonance scanning tunneling microscopy, we investigate coupled pairs of S=1/2 molecules (Fe–FePc), where one molecule serves as a readout and pumping unit for the neighboring one. We show that the exchange interaction between them enables angular momentum transfer, which allows for the control of the remote spin state via the direction and magnitude of the spin-polarized tunneling current and the exchange coupling strength. These results establish a general, all-electrical approach for remote spin initialization that is readily transferable to a wide range of spin-based quantum architectures.} 

\end{abstract}
\newpage

\section*{Introduction}
The controlled initialization of quantum states is a central requirement for quantum technologies\cite{divincenzo_physical_2000}, ranging from quantum computing to nanoscale sensing and simulation. At thermal equilibrium, however, the occupation of quantum states follows the Boltzmann distribution, which fundamentally limits the initialization fidelity by the ratio of energy splitting to temperature.\\

\noindent In this regard, a variety of approaches to actively control the occupation of quantum states has been developed across quantum platforms. Examples are energy-selective tunneling into an electron reservoir for quantum dots\cite{elzerman_single-shot_2004} as well as optical pumping for solid-state color centers\cite{chakraborty_polarizing_2017} and trapped ions\cite{cirac_quantum_1995}. Optical pumping has also been proposed and is being explored as a strategy for initializing molecular spin qubits\cite{sfataftah_progress_2018, vasilenko_optically_2026}. The latter have attracted growing interest as quantum building blocks due to their chemical tunability, intrinsic multilevel structure, and potential to scale up through self-assembly\cite{atzori_second_2019, gaita-arino_molecular_2019, moreno-pineda_molecular_2018, chiesa_molecular_2024}. Yet, their controlled initialization, particularly at the level of individual molecules, remains a major challenge. Addressing this requires approaches that are efficient, fast, and compatible with device integration. In this context, electrical control is appealing, as it provides direct addressability and high-speed operation without the need for extended optical setups.\\

\noindent  In order to access and manipulate individual spins with atomic precision, scanning tunneling microscopy combined with electron spin resonance (ESR-STM) has emerged as a powerful platform\cite{baumann_electron_2015}. By combining atomic-scale spatial resolution alongside energy resolution in the MHz regime ESR-STM enables the detection and coherent control of spin transitions in individual atoms, molecules and solid-state defects\cite{baumann_electron_2015, yang_coherent_2019, wang_atomic-scale_2023, zhang_electron_2022-1,  willke_coherent_2021, kovarik_spin_2024, au-yeung_atomic-scale_2026}. 
Recent works have further demonstrated the controlled polarization of the nuclear spin of a single atom via the tunneling current and transfer of angular momentum\cite{yang_electrically_2018, veldman_coherent_2024, stolte_single-shot_2025}. This approach offers a promising route for the initialization of electron spins in molecules.\\

\noindent In this work, we demonstrate current-induced polarization of a remote electron spin in a magnetic molecule. Using pairs of Fe-FePc complexes, we realize a system in which one molecule is addressed by the STM tip and serves as a sensing and pumping unit to remotely change and probe the state occupation of a neighboring molecule. By controlling the magnitude and polarity of the spin-polarized tunneling current we drive the remote spin into a non-thermal steady state, including occupation inversion by transferring angular momentum from the magnetic STM tip. Finally, we demonstrate how this electrical pumping can be varied by tuning the exchange interaction between spins and discuss its broader implications for spin-qubit platforms.
\newpage

\section*{Results and Discussion}
\begin{figure}[!htbp]
  \centering
  \includegraphics[width=160mm]{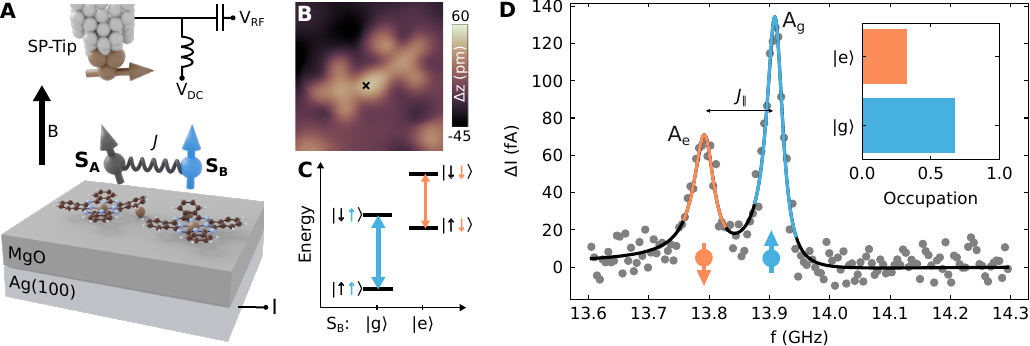} 
  \caption{\textbf{Measurement scheme.} (\textbf{A}) Schematic illustration of the tunnel junction with two molecule spins assembled on MgO/Ag(100). For ESR measurements an external magnetic field $B$ is applied and a magnetic tip apex is used. A DC bias voltage $V_\mathrm{DC}$ and a radio-frequency voltage $V_\mathrm{RF}$ are applied to the tip which is positioned above the sensing spin S$_\text{A}$ to probe and drive its spin state. S$_\text{A}$ is exchange coupled to a second, remote spin S$_\text{B}$. (\textbf{B}) STM topography of a molecule pair [Image conditions: $3~ \mathrm{nm}~\mathrm{x}~3~\mathrm{nm}, I=10~\mathrm{pA}, V_\mathrm{DC}=-100~\mathrm{mV}$]. The cross marks the measurement position used in (D). (\textbf{C}) Energy level diagram of the spin system, showing the configurations with the remote spin S$_\text{B}$ in the ground state $\ket{\mathrm{g}}$ (left) and the excited state $\ket{\mathrm{e}}$ (right). The two ESR transitions are indicated by colored arrows of different thickness. (\textbf{D}) ESR frequency sweep of the molecule pair in (B), showing two resonances associated with the two transitions shown in (C) corresponding to the two spin orientations of S$_\text{B}$. The peak amplitudes $A_\mathrm{g}$ and $A_\mathrm{e}$ are obtained from fitting a double Fano function as indicated. Inset: extracted state occupation of S$_\text{B}$ ($p_\mathrm{g}=67~\%, p_\mathrm{e} = 33~\%$). [ESR conditions: $G_\mathrm{set}=0.13~\mathrm{nS}, V_\mathrm{DC}=-50~\mathrm{mV}, B=458~\mathrm{mT}, V_\mathrm{RF}=10~\mathrm{mV}$].}
  \label{fig:fig1}
\end{figure}

\noindent \textbf{Investigated spin system}\\
The experimental platform consists of iron phthalocyanine (FePc) molecules and single iron (Fe) atoms adsorbed on two monolayers of magnesium oxide (MgO) on a Ag(100) crystal (see Fig. \ref{fig:fig1}A). The thin MgO film serves as a decoupling layer to protect the single spin hosts from substrate electrons while maintaining sufficient conductivity for a finite tunneling current to probe the surface spins\cite{paul_control_2017-1}. Using tip-assisted atom manipulation, single Fe atoms are combined with FePc molecules to form Fe-FePc complexes, where the additional Fe atom is positioned below one of the benzene rings of the FePc molecule\cite{huang_quantum_2025}. This coupled spin system exhibits an enhanced spin lifetime compared to pristine FePc molecules as shown in Ref.\cite{huang_quantum_2025}. It moreover forms an effective S=1/2 system and is treated in the following as such to simplify the discussion. Using these units as building blocks, molecule pairs are assembled by positioning two Fe-FePc complexes in close proximity with the STM tip as shown in Fig. \ref{fig:fig1}B. 
\\

\noindent For the following experiments, the magnetic STM tip, which is created by transferring several Fe atoms to the tip apex, is positioned above one Fe-FePc complex (sensing spin S$_{\text{A}}$). This spin then senses the neighboring Fe-FePc complex (remote spin S$_\text{B}$) via the exchange interaction between them. For that, we probe both spins in ESR measurements, where a radio-frequency (RF) voltage $V_\mathrm{RF}$ is applied to the STM tip. The spin readout is established by applying a DC bias voltage $V_\mathrm{DC}$ and measuring the spin-polarized tunneling current between magnetic tip and sample (see Fig. \ref{fig:fig1}A).\\
\newpage
\noindent \textbf{State occupation measurement}\\
To describe the coupled spin system of the two Fe–FePc complexes in greater detail, we employ an effective spin Hamiltonian\cite{yang_engineering_2017,zhang_electron_2022-1,veldman_free_2021,huang_quantum_2025} (see Supplementary Section \ref{sec:sup_spinhamil}):\\
\begin{equation}
H
=
g \mu_B \left( B + B_\mathrm{tip} \right) S_\mathrm{{A,z}}
+
g \mu_B B S_\mathrm{B,z} \\
+
J_\perp \left( S_\mathrm{{A,x}} S_\mathrm{{B,x}} + S_\mathrm{{A,y}} S_\mathrm{{B,y}} \right)
+
J_\parallel S_\mathrm{{A,z}} S_\mathrm{{B,z}}
\label{eq:2spinhamil}
\end{equation}\\
\noindent The first two terms describe the Zeeman energies of S$_\text{A}$ and S$_\text{B}$ with the gyromagnetic ratio $g\approx 2$\cite{huang_quantum_2025}. The local magnetic field $B_\mathrm{tip}$ originating from the magnetic tip allows to detune the two spins so that the eigenstates of the system are well described by the Zeeman product states of the two individual spins. The last two terms account for the interaction between S$_\text{A}$ and S$_\text{B}$, which is modeled here by an anisotropic exchange coupling with parallel and perpendicular component $J_{\parallel}$ and $J_{\perp}$ to the quantization axis, respectively. We note that the weaker dipole-dipole interaction is not separately considered for simplicity and is absorbed in the effective exchange parameters. \\
\noindent The resulting energy level diagram of the coupled spin system is shown in Fig. \ref{fig:fig1}C: Two ESR transitions are possible for S$_{\text{A}}$ that depend on the state of S$_{\text{B}}$\cite{choi_atomic-scale_2017}. \noindent Fig. \ref{fig:fig1}D shows a corresponding frequency sweep. The spectrum reveals two pronounced resonance frequencies separated by $\Delta f=|J_{\parallel}|$ (where we set Planck's constant $h=1$, see Supplementary Section \ref{sec:sup_spinhamil} for details). According to Fig. \ref{fig:fig1}C , the higher (lower) peak corresponds to the remote spin S$_\text{B}$ occupying the ground state $\ket{\mathrm{g}}$ (excited state $\ket{\mathrm{e}}$). Moreover, the amplitudes of the two resonances $A_\mathrm{g}$ and $A_\mathrm{e}$ are directly proportional to the state occupations $p_\mathrm{g},p_\mathrm{e}$ of S$_\text{B}$ following the relation $A_\mathrm{g}/A_\mathrm{e}=p_\mathrm{g}/p_\mathrm{e}$ (with $p_\mathrm{g}+p_\mathrm{e}=1$)\cite{choi_atomic-scale_2017}. Therefore, the ESR spectrum provides a direct measure of the state occupation of the remote spin.\\
\begin{figure}[!htbp]
  \centering
  \includegraphics[width=84mm]{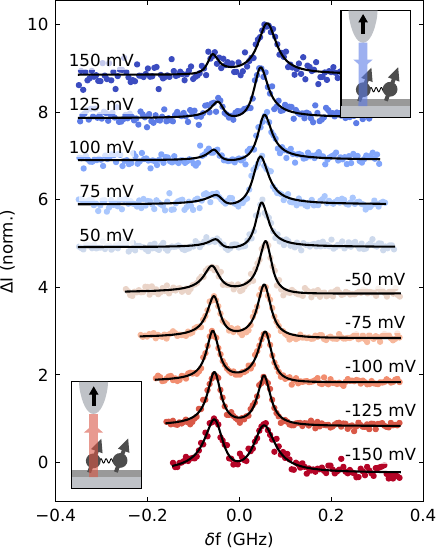}
  \caption{\textbf{Voltage dependence of the ESR spectra.} Change in tunneling current $\Delta I$ as a function of radio-frequency shift $\delta f=f-f_0$ for different $V_\mathrm{DC}$ at constant conductance setpoint $G_\mathrm{set}$ [ESR conditions: same as in Fig. \ref{fig:fig1}D]. The traces are normalized and vertically offset for clarity. In addition, they are centered at their resonance frequencies to compensate for the voltage induced frequency shift arising from spin-electric coupling\cite{greule_spin-electric_2025}. Colored dots show the experimental data and black solid lines represent fits to a double Fano function. The insets illustrate the tunnel junction with the direction of the tunneling current for positive and negative $V_\mathrm{DC}$.}
  \label{fig:fig2}
\end{figure}

\begin{figure}[!htbp]
  \centering
  \includegraphics[width=160mm]{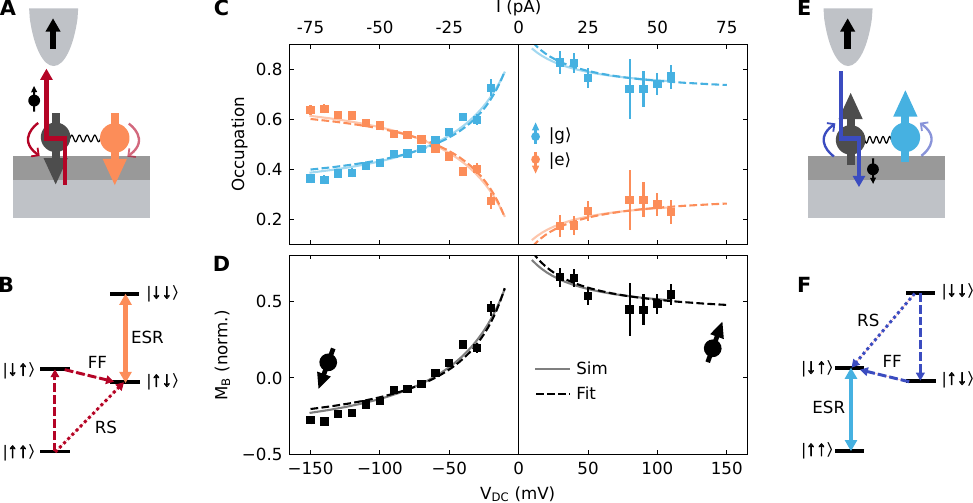}
  \caption{\textbf{Remote spin pumping in a molecule pair.} (\textbf{A}) Spin excitation: Schematic of the tunnel junction illustrating the remote spin excitation towards $\ket{\downarrow\downarrow}$ of the exchange coupled spins driven by the spin-polarized tunneling current for $I<0$. (\textbf{B}) Energy level diagram of the coupled spin system. Flip-flop (FF) and remote scattering (RS) transitions that dominate for $V_\mathrm{DC}<0$ are indicated by red arrows. ESR detection is shown in orange. (\textbf{C}) Extracted occupations of the ground $\ket{\mathrm{g}}$ and excited state $\ket{\mathrm{e}}$ of S$_\text{B}$ as a function of $V_\mathrm{DC}$ and $I$ at constant $G_\mathrm{set}$. Data points are obtained from ESR frequency sweeps as presented in Fig. \ref{fig:fig1}D [ESR conditions: $G_\mathrm{set}=0.5~\mathrm{nS}, B=450~\mathrm{mT}, V_\mathrm{RF}=12~\mathrm{mV}$]. (\textbf{D}) Corresponding magnetization $M_\mathrm{B}$ plotted over $V_\mathrm{DC}$ and $I$. Solid lines in the sub figures (C) and (D) represent fits to the full rate simulation [Simulation parameters: $T=0.21\pm0.12~\mathrm{K}, G_{st}/G_{ss}=0.17\pm0.03, \eta=0.34\pm0.04$]. Dashed lines represent fits to the analytical formula given by Eq. \ref{eq:M_B}: $I_0=-30.2\pm1.8~\mathrm{pA}, \eta=0.39\pm0.02$. For negative (positive) values $\ket{\mathrm{e}}$ ($\ket{\mathrm{g}}$) is primarily occupied as illustrated by the inserted sketches. (\textbf{E}) Spin relaxation: Schematic of the tunnel junction for $I>0$ illustrating the remote spin relaxation into $\ket{\uparrow\uparrow}$. (\textbf{F}) Energy level diagram with the FF and RS transitions that dominate for $V_\mathrm{DC}>0$ shown in blue and ESR detection in light blue.}
  \label{fig:fig3}
\end{figure}

\noindent\textbf{Electrically controlled spin polarization}\\
Fig. \ref{fig:fig2} presents a series of ESR frequency sweeps recorded at different $V_\mathrm{DC}$ at fixed tip height. Thus, the tunneling current $I=G_\mathrm{set}V_\mathrm{DC}$ increases where $G_\mathrm{set}$ is the constant conductance setpoint of the measurement.
For negative voltages, corresponding to electron tunneling from the sample to the tip, the relative amplitudes of the two ESR transitions change markedly: With decreasing $V_\mathrm{DC}$ the peak amplitudes become progressively more equal in height, eventually leading to a regime where the transition associated with S$_\text{B}$ being in the excited state $\ket{\mathrm{e}}$ becomes more dominant. 
In contrast, applying a positive bias voltage (current from tip to sample) has only a minor influence on the relative peak amplitudes. In Fig. \ref{fig:fig3}C we demonstrate this effect again by plotting obtained occupations $p_\mathrm{g}$ and $p_\mathrm{e}$ as a function of $V_\mathrm{DC}$ and $I$ (different dataset than Fig. \ref{fig:fig2}). Here, the occupations of the two states gradually approach for negative $V_\mathrm{DC}$ and eventually invert at $\approx-60~\mathrm{mV}$. In contrast, for positive $V_\mathrm{DC}$ the state occupations show only a weak dependence.
\\
From the state occupation of the remote spin we derive its magnetization as $M_\mathrm{B}=p_g-p_e$ (Fig. \ref{fig:fig3}D): $M_\mathrm{B}=1$ ($M_\mathrm{B}=-1$) corresponds to a full initialization in the ground (excited) state and thus $M_\mathrm{B}$ intuitively reflects the spin state initialization.\\
The observed occupation inversion for negative voltages cannot be reconciled with a thermal Boltzmann distribution of the spin system's state occupations as reported for other systems\cite{choi_atomic-scale_2017, del_castillo_theory_2025}. Further, the strong asymmetry with respect to the tunnel direction highlights the non-equilibrium nature of the remote spin state occupation and indicates a bias-polarity-dependent control mechanism. Moreover, by performing setpoint dependent measurements, we find that the spin pumping depends on the induced tunneling current $I$ instead of the explicit bias voltage $V_\mathrm{DC}$ (see Fig. \ref{fig:sup_setpoint}).\\

\newpage
\noindent \textbf{Inelastic Tunneling Theory}\\
To explain the observed change in state occupation, we employ a model of tunneling current induced spin pumping. This has been successfully applied to single atom nuclear spins\cite{yang_electrically_2018, stolte_single-shot_2025, veldman_coherent_2024}: In these experiments, the spin imbalance of the spin-polarized tip allows to polarize the electron spin via the tunneling current while the hyperfine interaction couples the electron and nuclear spin. The latter subsequently enables spin transfer between the two. Nevertheless, despite a large number of coupled spin systems reported in literature\cite{yang_engineering_2017,zhang_electron_2022-1,veldman_free_2021, yang_coherent_2019}, this has not been observed for electron or molecular spin systems.\\
\noindent To describe the influence of the tunneling current on the spin state occupation, we consider second-order tunneling processes between the spin system and the electrodes, following the framework introduced in Ref.\cite{loth_controlling_2010-1, ternes_spin_2015, loth_spin-polarized_2010, delgado_spin-transfer_2010}. In this approach the surface spins are treated as an open quantum system that is weakly coupled to the electronic reservoirs formed by the STM tip and the substrate\cite{wolf_-surface_2024}. Inelastic tunneling events induce transitions between the spin eigenstates and thereby redistribute the state occupations. In the tunnel junction geometry used in this work, the sensing spin S$_\text{A}$ is subject to inelastic scattering from both the tip and the sample, whereas the remote spin S$_\text{B}$ is only subject to electrons originating from the sample electrode. The magnetic tip electrode is modeled as a spin-polarized electron reservoir characterized by the polarization parameter $\eta$ which quantifies the imbalance between spin up (down) electron densities $n^t_\uparrow=\frac{1+\eta}{2}$ ($n^t_\downarrow=\frac{1-\eta}{2}$). In contrast, the sample electrode has equal electron densities $n^s_\uparrow = n^s_\downarrow=\frac{1}{2}$. Based on these tunneling processes and their rates, the state occupation of the spin system can be obtained numerically from the steady state solution of the corresponding rate equation (see Supplementary Section \ref{sec:sup_rates}). \\

\noindent To additionally establish an intuitive picture, we derive an analytical solution in the following: At sufficiently large bias voltages, $eV_\mathrm{DC}>E_i>k_\mathrm{B}T$, with $E_i$ being the eigenenergies of the spin system, inelastic scattering with tunneling electrons dominates over thermally activated transitions. This drives the spin system out of thermal equilibrium and enables current induced spin pumping. Here, the spin polarization of the tip causes an asymmetry between relaxation and excitation rates. For a given tip polarization $\eta$ the direction of $I$ determines whether relaxation or excitation is enhanced. This is illustrated in Fig. \ref{fig:fig3}A (E), where for $I<0$ ($I>0$) the surface spin is mainly excited (relaxed) by the tunneling electrons. This process competes with scattering processes in which electrons originate and return to the same electrode, namely tip-tip and sample-sample scattering: They drive the system towards thermal equilibrium. However, for S$_\text{A}$ these processes are negligible due to the dominant tunneling current, whereas for S$_\text{B}$ relaxation via substrate electrons (sample-sample scattering) remains a dominant process. \\   
\\
Although the tunneling electrons couple directly only to the sensing spin S$_\text{A}$, a finite state mixing between S$_\text{A}$ and S$_\text{B}$ enables the spin transfer to the remote spin. For that, we identify two contributing tunneling processes (Fig. \ref{fig:fig3}B,F): (i) excitation of S$_\text{A}$ is followed by a flip-flop interaction (FF) with the remote spin and (ii) a tunneling electron directly induces a transition of the remote spin via the small but finite state mixing (referred to as remote scattering RS). We can quantify the relative probabilities of this processes by computing the transition matrix elements of the corresponding spin states:\\ 
\begin{equation}
    \mathcal{M}_\mathrm{FF} = \frac{J_\perp^2}{4(\delta^2+J_\perp^2)}
\label{eq:RateFF}
\end{equation}

\begin{equation}
    \mathcal{M}_\mathrm{RS} =  \frac{1}{2}-\frac{\delta}{2\sqrt{\delta^2+J_\perp^2}}
\label{eq:RateRS}
\end{equation}\\
\noindent They depend on $J_\perp$ and the detuning $\delta=g \mu_\mathrm{B} B_\mathrm{tip}$, i.e. the difference in Zeeman energy of S$_\text{A}$ and S$_\text{B}$ resulting from the magnetic tip field $B_\mathrm{tip}$\cite{zhang_electron_2022-1}. Both parameters can be obtained experimentally as shown in Fig. \ref{fig:sup_jextraction}.
\noindent Based on the dominant processes described above we derive an analytical expression for the remote spin magnetization $M_\mathrm{B}$ as a function of the tunneling current $I$ (see Supplementary Section \ref{sec:sup_anasol} for details):\\
\begin{equation}
    M_\mathrm{B}(I) = \eta \frac{I-I_0}{|I|+|\eta I_0|}~~\text{with}~~I_0=-\frac{G_{ss} g \mu_\mathrm{B} B}{2e\mathcal{M} \eta}
\label{eq:M_B}
\end{equation}\\
\noindent where $I_0$ is the characteristic tunneling current at which $M_\mathrm{B}(I_0)=0$, i.e. how effectively the spin system is driven towards occupation inversion $p_\mathrm{e}=p_\mathrm{g}$. This process is facilitated by large flip-flop and remote scattering probabilities $\mathcal{M}=\mathcal{M}_\mathrm{FF}+\mathcal{M}_\mathrm{RS}$ and large spin polarization of the tip $\eta$. In contrast, spin relaxation via the surface, which is proportional to the conductance $G_{ss}$ and the Zeeman energy $g \mu_\mathrm{B} B$, hinders the spin pumping. In the limit of large tunneling currents ($I\gg I_0$), $M_\mathrm{B}$ approaches a steady state determined solely by the tip polarization $\eta$ [$M_\mathrm{B}\rightarrow -\eta$ for $V_\mathrm{DC}<0$ and $M_\mathrm{B}\rightarrow \eta$ for $V_\mathrm{DC}>0$]. Consequently, the achievable degree of spin initialization is limited by the spin polarization of the tip, which was found to lie in the range $0.2 \le \eta \le 0.5$ for the measurements presented here.\\

\noindent In Fig.~\ref{fig:fig3}C,D we plot the analytical expression (Eq. \ref{eq:M_B}) and the numerical solution of the rate equation model along with the experimental data and show excellent agreement. For the discussed dataset, we find a positive tip-polarization $\eta>0$ which favours excitations when tunneling from sample to tip, while also the opposite behavior can be obtained for magnetic tips with $\eta<0$ (see Fig. \ref{fig:sup_polarization}). \\

\begin{figure}[t]
  \centering
  \includegraphics[width=160mm]{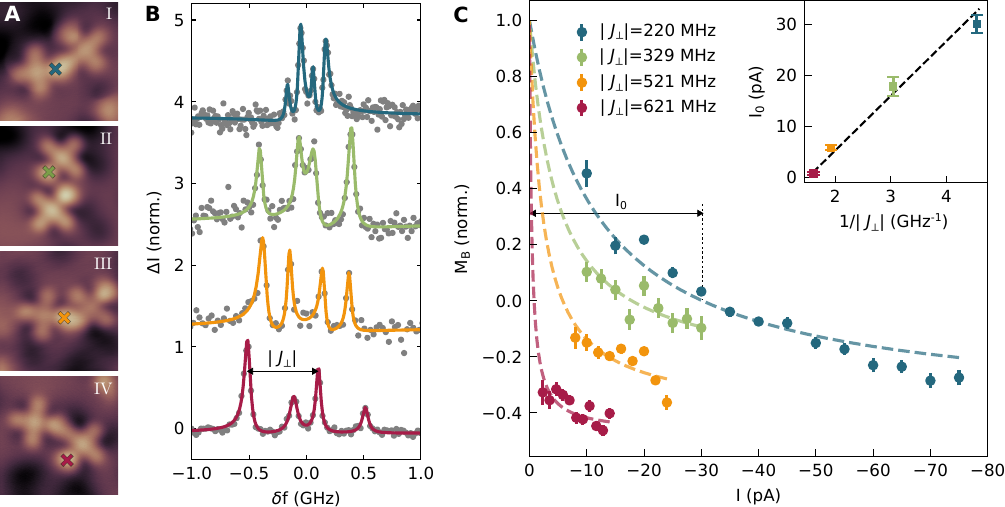}
  \caption{\textbf{Spin pumping for different exchange interaction strengths.} (\textbf{A}) STM topographies of four molecule pairs ($\mathrm{I-IV}$). Colored crosses mark the measurement positions used in B and C [Image conditions: $3~\mathrm{nm}~\mathrm{x}~3~\mathrm{nm},I=10\text{-}20~\mathrm{pA}, V_\mathrm{DC}=\pm 100~\mathrm{mV}$]. (\textbf{B}) Corresponding ESR spectra recorded at the avoided level crossing ($\delta =0$) showing $\Delta I$ as a function of radio frequency shift $\delta f$. Grey dots represent experimental data, solid lines are fits using a four peak Fano function. For clarity, the spectra are vertically offset. Four resonance peaks are resolved from which the perpendicular exchange interactions $|J_\perp|=f_3-f_1$ are extracted as indicated for pair $\mathrm{IV}$ (Also see Supplementary Section \ref{sec:sup_spinhamil} and Table \ref{tab:sup_parameters}). [ESR conditions: $G_\mathrm{set}=0.30\text{-}0.68~\mathrm{nS}, |V_\mathrm{DC}|=40\text{-}60~\mathrm{mV}, B=473\text{-}495~\mathrm{mT}, V_\mathrm{RF}=10\text{-}12~\mathrm{mV}$]. (\textbf{C}) Remote spin magnetization $M_\mathrm{B}$ as a function of the induced tunneling current $I$ for the four pairs. Dashed lines are fits to the experimental data (circles) based on Eq. \ref{eq:M_B} [ESR conditions: $G_\mathrm{set}=0.3\text{-}0.5~\mathrm{nS}, B=450\text{-}495~\mathrm{mT}, V_\mathrm{RF}=12~\mathrm{mV}$]. Stronger exchange coupling leads to an enhanced spin pumping efficiency, reflected by a reduced characteristic current $I_0$ which is obtained from the fits (indicated for pair $\mathrm{I}$). The inset displays $I_0$ plotted over $1/|J_\perp|$ with a linear fit (dashed line) as a guide to the eye.}
  \label{fig:fig4}
\end{figure}
\noindent \textbf{Tuning the exchange interaction} \\
The model we established in Eq. \ref{eq:M_B} illustrates that $M_\mathrm{B}$ depends on the transition matrix elements for inelastic scattering processes between S$_\text{A}$ and S$_\text{B}$,  $\mathcal{M}_\mathrm{RS}$ and $\mathcal{M}_\mathrm{FF}$. As both depend on the degree of state mixing, we expect a direct dependence on the perpendicular exchange interaction $J_\perp$ (Eq. \ref{eq:RateFF}, \ref{eq:RateRS}). Therefore, we construct pairs of Fe-FePc complexes with different spacing and orientation by tip-assisted atom manipulation (Fig. \ref{fig:fig4}A), which results in exchange couplings on the order of a few $100~\mathrm{MHz}$. The value $|J_\perp|$ for each pair is determined from ESR frequency sweeps recorded at the avoided level crossing ($\delta=0$) as shown in Fig. \ref{fig:fig4}B (see Fig. \ref{fig:sup_jextraction} for details). Fig. \ref{fig:fig4}C shows the magnetization $M_B(I)$ for all four complex pairs in the detuned regime $\delta>0$ analogous to Fig. \ref{fig:fig3}D.   
We find that pairs with stronger exchange coupling exhibit a significantly enhanced spin pumping efficiency, characterized by a faster approach to saturation with increasing $I$. This trend is quantitatively captured by the fitted characteristic tunneling current $I_0$: We find $I_0 \propto|1/J_\perp|$ (Fig. \ref{fig:fig4}C, inset) for the given parameter range in good agreement with Eqs.~\ref{eq:RateFF}--\ref{eq:M_B}, where a larger $|J_\perp|$ increases the degree of state mixing and thereby enhances the remote excitation and flip-flop process ($\mathcal{M}\propto {J_\perp}$ for $J_\perp<\delta$). As a consequence, engineering and tuning the exchange coupling provides an additional strategy for controlling the spin pumping, complementary to the voltage controlled tunneling current.\\
\noindent From the relation between $I_0$ and $\mathcal{M}(J_\perp)$ we find an estimate for the remote spin's lifetime $T_1 \approx 1\text{-}10~\mathrm{\mu s}$ (see Supplementary Section \ref{sec:sup_lifetime} for details). This agrees well with previous studies on the Fe-FePc complex\cite{huang_quantum_2025} and exceeds relaxation times reported for conventional S=1/2 systems on MgO/Ag(100)\cite{yang_coherent_2019, willke_coherent_2021}. We attribute the enhanced lifetime to the absence of tunneling through the remote spin and to the correlated ground state of the Fe-FePc complex\cite{huang_quantum_2025}. Importantly, this highlights another crucial condition for electric spin pumping, as the spin lifetime must be sufficiently long to enable state initialization before relaxation occurs.\\\\

\section*{Conclusion}
In this work, we demonstrate an  all-electrical route to remotely initialize (molecular) electron spins. Using exchange-coupled pairs of Fe-FePc complexes, we show that a spin-polarized tunneling current acting on one molecule can control the spin state occupation of the second, spatially separated molecule via transfer of angular momentum. The resulting non-equilibrium steady state is governed by the direction and magnitude of the tunneling current as well as the degree of its spin polarization.\\ 
Our results identify three essential ingredients required for electric spin pumping: (i) a magnetic electrode is necessary to provide a spin-polarized current which causes an imbalance between excitation and relaxation. (ii) In addition, a finite state mixing between the two molecular spins is required to mediate the angular momentum transfer to the remote spin which proceeds via two mechanisms: The first involves an initial pumping of the sensing spin followed by a flip-flop interaction with the remote spin. The second mechanism corresponds to a direct excitation of the remote spin. Both mechanisms can be efficiently tuned by the strength of the perpendicular exchange between the spins.
(iii) Eventually, these inelastic tunneling rates must exceed the intrinsic relaxation to drive the system out of thermal equilibrium.\\
Our analysis shows that the molecular spins will eventually adapt the spin polarization of the tip. While this sets a limit to their effective initialization, we believe that an improved spin polarization close to $100\%$ can be achieved via tip functionalization with magnetic molecules\cite{bae_direct_2024} or Yu–Shiba–Rusinov states\cite{eltschka_probing_2014}.\\
Beyond the specific molecular spin system studied here, our findings demonstrate  how remote spin pumping of individual electron spins can be achieved generally via electric currents. Such electric control provides a viable route for other implementation into quantum architectures such as quantum dots or solid state spin defects. The ability to induce spin polarization beyond thermal equilibrium further allows in principle to perform single spin ESR at higher temperatures (see Fig. \ref{fig:sup_temperature}) as realized in optically detected magnetic resonance\cite{gruber_scanning_1997}. In addition, electron spin pumping can be extended to spin chains (see Fig. \ref{fig:sup_3spinchain}) which highlights its potential for initializing larger qubit arrays. These capabilities are essential for scaling toward whole molecular spin networks, quantum simulators, and distributed quantum architectures. 

\section*{Acknowledgements}
P.W. acknowledges funding from the Emmy Noether Programme of the Deutsche  Forschungsgemeinschaft (DFG, WI5486/1-2). P.G. and P.W. acknowledge financial support from the Hector Fellow Academy (Grant No. 700001123). P.W. and K.H.A.Y. acknowledge support from the Center for Integrated Quantum Science and Technology (IQST). P.W. and J.S. acknowledge funding from the ERC Starting Grant ATOMQUANT. P.G. and P.W. thank Markus Ternes and Amogh Kinikar for valuable discussions. 

\section*{Author contributions}
P.G., W.H., K.H.A.Y and P.W. conceived the experiment. P.G., W.H., M.S., K.H.A.Y., J.S., C.S., W.W. and P.W. set up the experiment and conducted the measurements. P.G., W.H., K.H.A.Y. and P.W. analyzed and discussed the experimental data. P.G. performed the rate simulations. P.G. and P.W. wrote the manuscript with input from all authors. W.W. and P.W. supervised the project.

\section*{Data availability}
The data that support the findings of this study are available from the corresponding author upon reasonable request.

\printbibliography[title={References}]

\newpage

\clearpage
\renewcommand{\thefigure}{S\arabic{figure}}
\renewcommand{\thetable}{S\arabic{table}}
\renewcommand{\theequation}{S\arabic{equation}}
\renewcommand{\thepage}{S\arabic{page}}
\renewcommand{\thesection}{S\arabic{section}}
\setcounter{figure}{0}
\setcounter{table}{0}
\setcounter{equation}{0}
\setcounter{page}{1} 

\begin{center}
\section*{Supplementary Material for\\ \papertitle}
\vspace{0.5cm}
\AuthorList
\end{center}
\vspace{0.5cm}
\Affiliations \\
\CorrespondingAuthor

\tableofcontents
\clearpage

\section{Methods}

\textbf{Experimental Setup}\\
The measurements were conducted in a Unisoku USM1600 scanning tunneling microscope mounted on a home-built dilution refrigerator operating at a base temperature of $50~\mathrm{mK}$. The DC bias voltage $V_\mathrm{DC}$ and radio frequency voltage $V_\mathrm{RF}$ were applied to the tip. For the data evaluation the sign of the stated $V_\mathrm{DC}$ values was changed corresponding to the convention with bias applied to the sample. The microwave signal was generated by a Rohde \& Schwarz SMB100B generator and mixed with $V_\mathrm{DC}$ through a Microwave MDPX-0305 Diplexer from Marki. For the ESR measurements, magnetic tips were prepared by picking up several Fe atoms from the MgO surface. The used tips usually showed a step at $V_\mathrm{DC}=0~\mathrm{mV}$ in differential conductance (dI/dV) measurements at the Fe site of the Fe-FePc molecules stemming from inelastic tunneling with a spin polarized tip \cite{ternes_spin_2015}. In addition, a magnetic field perpendicular to the sample surface was applied. For ESR frequency sweeps we used a lock-in detection scheme where the RF voltage $V_\mathrm{RF}$ was modulated at 323 Hz while the DC voltage $V_\mathrm{DC}$ was applied continuously. For read out we used a Stanford Research Systems SR860 digital lock-in amplifier.\\     
\\
\textbf{Sample Preparation}\\
The sample was prepared in-situ under ultra-high vacuum with a base pressure of $<5\mathrm{x}10^{-10}~\mathrm{mbar}$. The Ag(100) crystal was cleaned by multiple cycles of argon ion sputtering and subsequent annealing. Afterwards, magnesium oxide islands were grown by evaporating Mg in an oxygen-rich atmosphere for $10~\mathrm{min}$ onto the heated silver crystal held at $510~^{\circ}  \mathrm{C}$ with an oxygen pressure of ca. $1\mathrm{x}10^{-6}~\mathrm{mbar}$. The deposition resulted in a coverage of ca. $50~\%$ with island thicknesses ranging from two to five monolayers. Via a home-built Knudsen cell FePc molecules were deposited onto the surface with a deposition time of ca. $90~\mathrm{s}$. For the subsequent Fe atom deposition the sample was cooled down to ca. $10-20~\mathrm{K}$. The Fe atoms were deposited for $21~\mathrm{s}$ using e-beam evaporation leading to a sub-monolayer coverage.

\newpage
\section{Spin Hamiltonian}
\label{sec:sup_spinhamil}

\noindent We present here the spin Hamiltonian used in this work together with its solution as it forms the basis for deriving the transition matrix elements discussed in section \ref{sec:sup_anasol} below as well as the experimental determination of the spin Hamiltonian parameters. As outlined in Ref.\cite{huang_quantum_2025} Fe-FePc complexes can be modeled as effective S=1/2 systems. We therefore describe the molecule pairs consisting of two Fe-FePc complexes with the following Hamiltonian\cite{yang_engineering_2017,zhang_electron_2022-1,veldman_free_2021,huang_quantum_2025} (Also see Eq. \ref{eq:2spinhamil}):\\
\begin{equation}
H
=
g \mu_B \left( B + B_{\mathrm{tip}} \right) S_\mathrm{{A,z}}
+
g \mu_B B S_\mathrm{{B,z}} \\
+
J_{\perp} \left( S_\mathrm{{A,x}} S_\mathrm{{B,x}} + S_\mathrm{{A,y}} S_\mathrm{{B,y}} \right)
+
J_{\parallel} S_\mathrm{{A,z}} S_\mathrm{{B,z}}
\label{eq:sup_2spinhamil}
\end{equation}\\
With the respective Zeeman terms where the sensing spin S$_\text{A}$ experiences an additional tip field $B_\mathrm{tip}$ to the external magnetic field $B$. The exchange interaction is parametrized with the perpendicular $J_\perp$ and parallel interaction $J_\parallel$ with respect to the quantization axis, given by the external magnetic field. We note, that the weaker dipole-dipole interaction is absorbed in the exchange parameters. $g$ is the g-factor and $\mu_\mathrm{B}$ the Bohr magneton. For $B>J_\parallel$ diagonalization of Eq.~\ref{eq:sup_2spinhamil} yields four eigenstates which can be expressed as: \\
\begin{align}
\ket{0} &= \ket{\uparrow\uparrow} \label{eq:eigen_1} \\
\ket{1} &= \frac{-\alpha}{\sqrt{1+\alpha^2}} \ket{\uparrow\downarrow}
        +  \frac{1}{\sqrt{1+\alpha^2}} \ket{\downarrow\uparrow} \label{eq:eigen_2} \\
\ket{2} &= \frac{1}{\sqrt{1+\alpha^2}} \ket{\uparrow\downarrow}
        +  \frac{\alpha}{\sqrt{1+\alpha^2}} \ket{\downarrow\uparrow} \label{eq:eigen_3} \\
\ket{3} &= \ket{\downarrow\downarrow} \label{eq:eigen_4}
\end{align}\\
With the parameter $\alpha=\frac{\delta+\sqrt{\delta^2 + J_{\perp}^2}}{J_{\perp}}$ and the detuning $\delta=g \mu_{\mathrm{B}} B_{\mathrm{tip}}$ quantifying the difference of the Zeeman energies of the two individual spins. The eigenenergies of the spin system are given by: 
\begin{align}
E_0 &= \frac{1}{4}\left(-2 g \mu_{\mathrm{B}} B_{\mathrm{tip}} - 4 g \mu_{\mathrm{B}} B + J_{\parallel}\right) \label{eq:eigenE_1} \\[6pt]
E_1 &= \frac{1}{4}\left(-J_{\parallel} - 2 \sqrt{\delta^2 + J_{\perp}^2}\right) \label{eq:eigenE_2} \\[6pt]
E_2 &= \frac{1}{4}\left(-J_{\parallel} + 2 \sqrt{\delta^2 + J_{\perp}^2}\right) \label{eq:eigenE_3} \\[6pt]
E_3 &= \frac{1}{4}\left(2 g \mu_\mathrm{B} B_{\mathrm{tip}} + 4 g \mu_\mathrm{B} B + J_{\parallel}\right) \label{eq:eigenE_4}
\end{align}
\begin{figure}[!htbp]
  \centering
  \includegraphics[width=155mm]{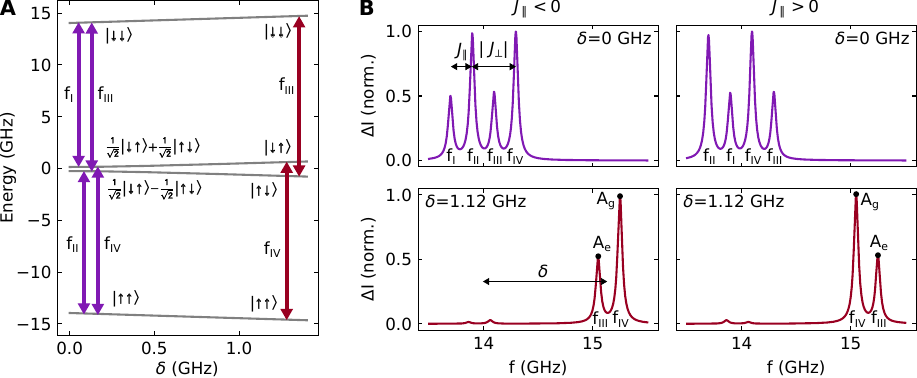}
  \caption{\textbf{Simulated ESR spectra.} (\textbf{A}) Energy level diagram of the coupled spin system as a function of detuning $\delta$. Violet arrows illustrate the four possible transitions at the avoided level crossing $\delta=0$. Red arrows illustrate the two remaining transitions in the detuned regime ($\delta>0$). (\textbf{B}) ESR simulations ($\Delta I$ over $f$) for ferromagnetic (left) and antiferromagnetic (right) exchange coupling in the two regimes ($\delta=0, \delta>0$). Through the resonance peaks the exchange interaction parameters $|J_\perp|$ and $J_\parallel$, the state occupations $p_\mathrm{g}\propto A_\mathrm{g}, p_\mathrm{e}\propto A_\mathrm{e}$ and detuning $\delta$ can be obtained. [Simulation parameters: $B=500~\mathrm{mT}, g=2, |J_\perp|=400~\mathrm{MHz}, J_\parallel=200~\mathrm{MHz}, T=1~\mathrm{K}$]}
  \label{fig:sup_HTheo}
\end{figure}

\noindent A corresponding energy level diagram is displayed in Fig. \ref{fig:sup_HTheo}A. Generally, four ESR transitions with $\Delta m=\pm1$ are allowed: $f_{\mathrm{I}}=E_3-E_2$, $f_{\mathrm{II}}=E_1-E_0$, $f_{\mathrm{III}}=E_3-E_1$ and $f_{\mathrm{IV}}=E_2-E_0$. We note that we put the Planck constant $h=1$. The resonance frequencies read:\\
\begin{align}
f_{\mathrm{I}} &= \frac{1}{2} \left( 2 g \mu_{\mathrm{B}} B+g \mu_{\mathrm{B}} B_{\mathrm{tip}} +  J_\parallel - \sqrt{\delta^2+J_\perp^2}  \right)\\[6pt]
f_{\mathrm{II}} &=  \frac{1}{2} \left( 2 g \mu_{\mathrm{B}} B+g \mu_{\mathrm{B}} B_{\mathrm{tip}} -  J_\parallel - \sqrt{\delta^2+J_\perp^2}  \right)\\[6pt]
f_{\mathrm{III}} &=  \frac{1}{2} \left( 2 g \mu_{\mathrm{B}} B+g \mu_{\mathrm{B}} B_{\mathrm{tip}} +  J_\parallel + \sqrt{\delta^2+J_\perp^2}  \right)\\[6pt]
f_{\mathrm{IV}} &=  \frac{1}{2} \left( 2 g \mu_{\mathrm{B}} B+g \mu_{\mathrm{B}} B_{\mathrm{tip}} -  J_\parallel + \sqrt{\delta^2+J_\perp^2}  \right)
\end{align}

\begin{figure}[!htbp]
  \centering
  \includegraphics[width=0.9\linewidth]{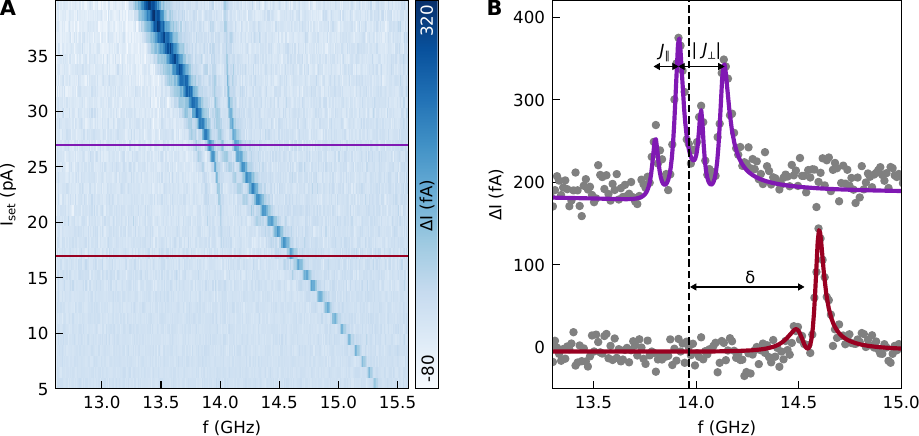}
  \caption{\textbf{Extraction of coupling parameters.} (\textbf{A}) Colormap of a ESR frequency sweep measured on a ferromagnetically coupled molecule pair showing the change in tunneling current $\Delta I$ as a function of radio frequency $f$ and setpoint current $I_\mathrm{set}$. As the setpoint is changed, the systems energy levels are tuned by the tip magnetic field through the avoided level crossing at which $\delta=0$. The spectrum evolves from two to four and back to two resonance peaks, reflecting the hybridization of the spins close to $\delta=0$. Solid horizontal lines indicate the traces displayed in (B) [ESR conditions: $V_\mathrm{DC}=40~\mathrm{mV}, B=494~\mathrm{mT}, V_\mathrm{RF}=10~\mathrm{mV}$]. (\textbf{B}) Single spectra showing $\Delta I$ over $f$. The upper trace (violet) shows the ESR signal measured at $\delta=0$ from which the in-plane and out-of-plane exchange interactions,  $|J_\perp|$ and $J_\parallel$, are extracted (black arrows). The lower trace (red) exemplarily shows a frequency sweep in the detuned regime. The detuning $\delta$ is determined from the frequency shift relative to the avoided level crossing. The traces are vertically offset for clarity.}
  \label{fig:sup_jextraction}
\end{figure}

\noindent \textbf{Zero Detuning}\\
For $\delta=0$ all four transitions ($f_\mathrm{I}\text{-}f_\mathrm{IV}$) are allowed when performing ESR on the sensing spin S$_\text{A}$ caused by the state mixing (see Fig. \ref{fig:sup_HTheo}B top). From the resonance frequencies the exchange interaction parameters can be determined: $|J_\perp|=f_\mathrm{III}-f_\mathrm{I}=f_\mathrm{IV}-f_\mathrm{II}$ and $J_\parallel=f_\mathrm{I}-f_\mathrm{II}=f_\mathrm{III}-f_\mathrm{IV}$. The sign of $J_\parallel$ further reveals whether the exchange is ferromagnetic $J_\parallel<0$ or antiferromagnetic $J_\parallel>0$. Here, one must be able to assign the observed resonance peaks to the respective transitions. If the systems state occupation is governed by the thermal Boltzmann distribution, this can be achieved by the relative peak amplitudes. For ferromagnetic coupling the resonances corresponding to $f_\mathrm{I}$ and $f_\mathrm{III}$ are less pronounced than $f_\mathrm{II}$ and $f_\mathrm{IV}$ because they involve eigenstates with higher energy which are less occupied. In contrast, for antiferromagnetic coupling $f_\mathrm{I}$ and $f_\mathrm{III}$ are more pronounced than $f_\mathrm{II}$ and $f_\mathrm{IV}$ (see Fig. \ref{fig:sup_HTheo}B). These relations allow to determine the coupling type and the magnitude of $J_\perp$ and $J_\parallel$. Table \ref{tab:sup_parameters} presents the determined parameters for the complex pairs presented in Fig. \ref{fig:fig4}. We note that in order to provide a solid analysis one must be careful with the assignment of the resonance peaks when spin pumping is present. In this case one must operate at low currents or in the regime where spin relaxation is enhanced to assure that the state occupations are not inverted. \\        
\\
\noindent \textbf{Detuned Regime}\\
For the detuned regime, where $B_{\mathrm{tip}}>0$ and thus $\delta>0$ as well as $|B_{\mathrm{tip}}|>|J_\perp|$, the eigenstates are well presented by the Zeeman product states having $\ket{1} \approx \ket{\uparrow\downarrow}$ and $\ket{2} \approx \ket{\downarrow\uparrow}$. In this regime the voltage-dependent frequency sweeps presented in this work are performed. Here, only the two transitions corresponding to $f_\mathrm{III}$ and $f_\mathrm{IV}$ are allowed (see Fig. \ref{fig:sup_HTheo}B bottom). The detuning $\delta$ is obtained from the relative frequency shift with respect to the avoided level crossing. An experimental example is shown in Fig. \ref{fig:sup_jextraction}. 
\noindent Knowing the coupling type from the evaluation strategy described above we can determine the state occupation. For ferromagnetic coupling $J_\parallel<0$ we find the occupation of the remote spin's ground state $\ket{\mathrm{g}}$ to correspond to the resonance peak at higher frequency. For the antiferromagnetic case $J_\parallel>0$ we find the occupation of $\ket{\mathrm{g}}$ to correspond to the resonance peak at lower frequency.\\ 
  


\begin{table}[!htbp]
\centering
\begin{tabular}{c c c c c}
\hline
 Molecule Pair No. & $\mathrm{I}$  & $\mathrm{II}$ & $\mathrm{III}$ & $\mathrm{IV}$  \\
\hline
 $|J_\perp| ~(\mathrm{MHz})$ & $220\pm3.4$ & $329\pm4.5$ & $521\pm4.9$ & $621\pm2.9$ \\
 $J_\parallel ~(\mathrm{MHz})$ & $-135\pm1.8$ & $457\pm4.6$ & $-220\pm1.0$ & $-403\pm1.2$ \\
 $\delta~(\mathrm{MHz})$ & $2049\pm1.7$ & $769\pm2.3$ & $790\pm2.6$ & $791\pm1.5$ \\
\hline
\end{tabular}
\caption{\textbf{Coupling Parameters.} Summary of the exchange interaction parameters $|J_\perp|,J_\parallel$ and the detuning $\delta$ determined from the measurements shown in Fig. \ref{fig:fig4}B,C.}
\label{tab:sup_parameters}
\end{table}

\clearpage
\FloatBarrier
\section{Rate Simulation}
\label{sec:sup_rates}

\begin{figure}[!htbp]
  \centering
  \includegraphics[width=80mm]{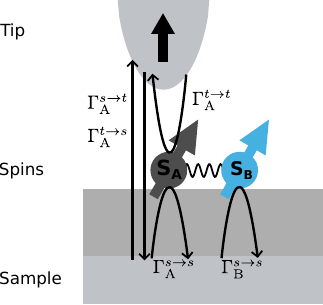}
  \caption{\textbf{Schematic of the tunnel junction geometry.} The image shows the tunneling processes considered in the full rate simulation. The surface spins are separated from the tip and sample electrodes by the vacuum barrier and the insulating MgO layer. Both spins undergo elastic and inelastic scattering with sample electrons. The sensing spin S$_\text{A}$, located below the spin-polarized tip, additionally couples to tip electrons. This enables additional tip-tip as well as tip-sample and sample-tip tunneling processes that allow the control over relaxation and excitation.}
  \label{fig:sup_rates}
\end{figure}

To model the state occupation of the spin system under the influence of the tunneling current we describe the interaction with the tunneling electrons via second order scattering rates $\Gamma$. As mentioned in the main text, we model the tip ($t$) and sample ($s$) electrodes by two electron reservoirs with $n^t_\uparrow=\frac{1+\eta}{2}$,  $n^t_\downarrow=\frac{1-\eta}{2}$ and $n^s_\uparrow = n^s_\downarrow=\frac{1}{2}$ with the spin polarization of the tip $\eta$. 
In our model, the sensing spin S$_\text{A}$ experiences tunneling rates from tip and sample $\Gamma^{s\rightarrow t}_\mathrm{A}$, $\Gamma^{t\rightarrow s}_\mathrm{A}$, $\Gamma^{t\rightarrow t}_\mathrm{A}$ and $\Gamma^{s\rightarrow s}_\mathrm{A}$. The remote spin S$_\text{B}$ which is distant from the tip electrode only experiences scattering with electrons from the surface $\Gamma^{s\rightarrow s}_\mathrm{B}$. A schematic drawing of the tunneling processes is shown in Fig. \ref{fig:sup_rates}. We note here, that the transition rates induced by the ESR detection are small compared to the excitation and relaxation rates from electron scattering\cite{yang_electrically_2018}.\\

\noindent According to Ref.\cite{ternes_spin_2015, loth_controlling_2010-1, delgado_spin-transfer_2010, loth_spin-polarized_2010} the tunneling rates from state $i$ to $f$  of the spin system are given by:

\begin{equation}
    \Gamma^{s\rightarrow t}_{\mathrm{A},if}=\frac{G_{st}}{e^2}P_{\mathrm{A},if}^{s\rightarrow t} \int_{-\infty}^{\infty} f(\epsilon+eV_\mathrm{DC},T)[1-f(\epsilon-E_{if},T)]d\epsilon 
\end{equation}

\begin{equation}
    \Gamma^{t\rightarrow s}_{\mathrm{A},if}=\frac{G_{st}}{e^2}P_{\mathrm{A},if}^{t\rightarrow s}\int_{-\infty}^{\infty} f(\epsilon-eV_\mathrm{DC},T)[1-f(\epsilon-E_{if},T)]d\epsilon 
\end{equation}

\begin{equation}
    \Gamma^{t\rightarrow t}_{\mathrm{A},if}=\frac{G_{tt}}{e^2}P_{\mathrm{A},if}^{t\rightarrow t} \int_{-\infty}^{\infty} f(\epsilon,T)[1-f(\epsilon-E_{if},T)]d\epsilon 
\end{equation}

\begin{equation}
    \Gamma^{s\rightarrow s}_{\mathrm{A},if}=\frac{G_{ss}}{e^2}P_{\mathrm{A},if}^{s\rightarrow s} \int_{-\infty}^{\infty} f(\epsilon,T)[1-f(\epsilon-E_{if},T)]d\epsilon 
\end{equation}

\begin{equation}
    \Gamma^{s\rightarrow s}_{\mathrm{B},if}=\frac{G_{ss}}{e^2}P_{\mathrm{B},if}^{s\rightarrow s} \int_{-\infty}^{\infty} f(\epsilon,T)[1-f(\epsilon-E_{if},T)]d\epsilon 
\end{equation}\\
\noindent Here, $e$ is the electron charge, $G_{st}$ the zero voltage conductance of the tunnel junction, $G_{ss}$ the sample-sample conductance and $G_{tt}$ the tip-tip conductance. $E_{if}=E_f-E_i$ is the energy difference of the involved states (see Eq. \ref{eq:eigenE_1}-\ref{eq:eigenE_4}). $P_{\mathrm{A},if}$ and $P_{\mathrm{B},if}$ are scattering probabilities based on the matrix elements between electrons with spin S$_\text{A}$ and spin S$_\text{B}$ as described in Ref.\cite{loth_controlling_2010-1}. 

\noindent The state occupation $p_i$ is then given by the steady state solution of the rate equation ($t\rightarrow\infty$):

\begin{equation}
    \frac{d}{dt}p_i(t)= \sum_{f \neq i} p_f(t)\Gamma_{fi}-p_i(t) \sum_{f \neq i} \Gamma_{if}
\label{eq:sup_rateequation}
\end{equation}

\noindent With $\Gamma_{if}=\Gamma^{t\rightarrow s}_{\mathrm{A},if} + \Gamma^{s\rightarrow t}_{\mathrm{A},if} + \Gamma^{t\rightarrow t}_{\mathrm{A},if} + \Gamma^{s\rightarrow s}_{\mathrm{A},if} + \Gamma^{s\rightarrow s}_{\mathrm{B},if}$ the sum of all rates. 

\clearpage
\section{Analytical Approximation}
\label{sec:sup_anasol}

To find an analytical solution of the rate equation (Eq. \ref{eq:sup_rateequation}) for the spin system described in section \ref{sec:sup_spinhamil} we first compute the state occupation of the sensor spin S$_\text{A}$ and then in a second step the occupation of the remote spin S$_\text{B}$. For computing the rates of the individual S=1/2 spins, we use the rate equation of a two-level system which reads as:\\

\begin{equation}   
\frac{d}{dt}
\begin{pmatrix}
p_{\mathrm{g}} \\
p_{\mathrm{e}}
\end{pmatrix}
=
\begin{pmatrix}
-\Gamma_{\mathrm{exc}} & \Gamma_{\mathrm{relax}} \\
\Gamma_{\mathrm{exc}} & -\Gamma_{\mathrm{relax}}
\end{pmatrix}
\begin{pmatrix}
p_\mathrm{g} \\
p_\mathrm{e}
\end{pmatrix}
\end{equation}\\

\begin{figure}[!htbp]
  \centering
  \includegraphics[width=150mm]{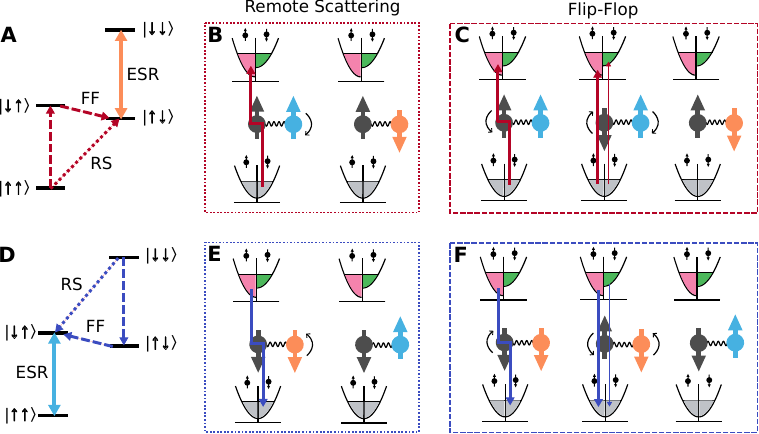}
  \caption{\textbf{Remote Scattering and Flip-Flop process.} (\textbf{A}) Energy level diagram with the remote scattering process (red dotted line), the flip-flop process (red dashed line) and the ESR detection (orange) indicated with arrows for current from sample to tip ($I<0$). (\textbf{B}) Schematic of the inelastic scattering in the tunnel junction for the remote scattering. The tip is modeled via a spin-polarized electron bath ($\eta>0$) with a higher density of states for up electrons (pink) than down electrons (green). In contrast, the sample electron bath is spin averaged. An electron tunnels from tip to sample transferring its angular momentum directly onto the remote spin causing it to be spin pumped into the excited state. (\textbf{C}) Schematic of the flip-flop interaction. First an electron excites the sensing spin, then a second electron mediates the flip-flop process between the sensing and remote spin. (\textbf{D-F}) The drawings show the respective processes discussed in (A-C) for the tunneling current flowing from tip to sample ($I>0$) which causes the remote spin to relax.}
  \label{fig:sup_RSandFF}
\end{figure}

\begin{figure}[!htbp]
  \centering
  \includegraphics[width=140mm]{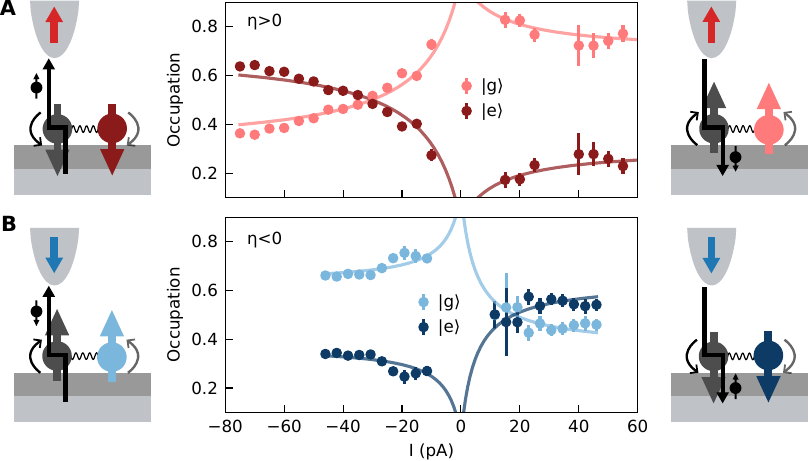}
  \caption{\textbf{State occupation for opposite tip polarizations.} (\textbf{A}) State occupations as a function of $I$ for a positive tip polarization $\eta>0$ [ESR conditions: $G_\mathrm{set}=0.5~\mathrm{nS}, B=450~\mathrm{mT}, V_\mathrm{RF}=12~\mathrm{mV}$]. The solid lines are fits to the experimental data (dots) based on Eq.~\ref{eq:M_B} [fit results: $\eta=0.38~\pm~0.02, I_0=-15.0~\pm~1.0~\mathrm{pA}$]. The sketch to the left (right) illustrates the spin excitation (relaxation) caused by the spin-polarized tunneling current for $\eta>0$. (\textbf{B}) State occupations as a function of $I$ for $\eta<0$ [ESR conditions: $G_\mathrm{set}=0.77~\mathrm{nS}, B=450~\mathrm{mT}, V_\mathrm{RF}=8~\mathrm{mV}$, fit results: $\eta=-0.27~\pm~0.02, I_0=18.1~\pm~1.7~\mathrm{pA}$]. 
  }
  \label{fig:sup_polarization}
\end{figure}

\begin{figure}[!htbp]
  \centering
  \includegraphics[width=140mm]{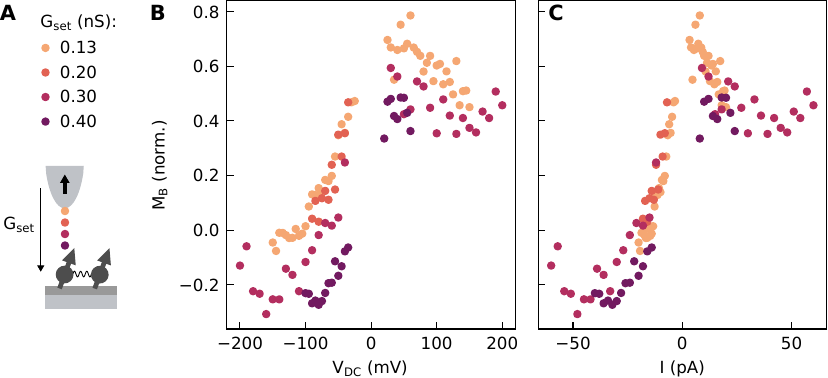}
  \caption{\textbf{Setpoint dependence.} (\textbf{A}) Schematic of the tunnel junction illustrating how the STM tip is positioned closer to the surface when increasing the setpoint conductance $G_\mathrm{set}=I_\mathrm{set}/V_\mathrm{set}$. (\textbf{B}) Magnetization of the remote spin $M_\mathrm{B}$ as a function of bias voltage $V_\mathrm{DC}$ for different $G_\mathrm{set}$ measured on the same molecule pair. Increasing $G_\mathrm{set}$ enhances the spin pumping efficiency, reflected by a steeper evolution of $M_\mathrm{B}$ with bias. [ESR conditions: $G_\mathrm{set}=0.13~\mathrm{nS}, B=458~\mathrm{mT}, V_\mathrm{RF}=10~\mathrm{mV}, G_\mathrm{set}=0.2~\mathrm{nS}, B=458~\mathrm{mT}, V_\mathrm{RF}=10~\mathrm{mV},G_\mathrm{set}=0.3~\mathrm{nS}, B=494~\mathrm{mT}, V_\mathrm{RF}=10~\mathrm{mV},G_\mathrm{set}=0.4~\mathrm{nS}, B=484~\mathrm{mT}, V_\mathrm{RF}=12~\mathrm{mV}$]. (\textbf{C}) Same data as in (B) showing $M_\mathrm{B}$ over $I=G_\mathrm{set}V_\mathrm{DC}$. The collapse of the traces onto a single curve demonstrates that spin pumping is governed by the induced current $I$ rather than the explicit applied bias voltage $V_\mathrm{DC}$.} 
  \label{fig:sup_setpoint}
\end{figure}

\noindent With the occupation of the ground (excited) state $p_\mathrm{g}$ ($p_\mathrm{e}$) and the sum of all rates that excite (relax) the spin $\Gamma_\mathrm{exc}$ ($\Gamma_\mathrm{relax}$). We find the steady state solution for $t\rightarrow \infty$: $p_\mathrm{g}=\frac{\Gamma_\mathrm{relax}}{\Gamma_\mathrm{exc}+\Gamma_\mathrm{relax}}$ and $p_\mathrm{e}=\frac{\Gamma_\mathrm{exc}}{\Gamma_\mathrm{exc}+\Gamma_\mathrm{relax}}$ and consequently the magnetization $M=\frac{\Gamma_\mathrm{relax}-\Gamma_\mathrm{exc}}{\Gamma_\mathrm{relax}+\Gamma_\mathrm{exc}}$.\\ 
\\
\noindent \textbf{Sensing Spin}\\
For S$_\text{A}$ the rates from tip to sample and sample to tip are dominant, since the potential difference is the dominant energy scale $eV_\mathrm{DC}>E_{if}>k_\mathrm{B} T$ while the conductances are of similar magnitude $G_{ss}\approx G_{tt} \approx G_{st}$. Therefore, we can neglect the inelastic transitions from surface-surface or tip-tip processes. We also neglect the finite terms that include scattering with the remote spin due to the low mixing of the spin states in the detuned regime. The excitation and relaxation rates of the sensing spin are then given by:\\
\begin{equation}
\label{eq:RatesA}
\Gamma_\mathrm{exc} =
\dfrac{G_{st}}{e}\, |V_\mathrm{DC}| \, \frac{1}{2}\frac{1 \mp \eta}{2}
\end{equation}
\begin{equation}
\label{eq:RatesA}
\Gamma_\mathrm{relax} =
\dfrac{G_{st}}{e}\, |V_\mathrm{DC}| \, \frac{1}{2}\frac{1 \pm \eta}{2}
\end{equation}\\
\noindent where the upper (lower) sign applies to $V_\mathrm{DC}>0$ ($V_\mathrm{DC}<0$). For $V_\mathrm{DC}>0$ we obtain $p_\mathrm{A,g}=\frac{1+\eta}{2}$, $M_\mathrm{A}=\eta$ and for $V_\mathrm{DC}<0$: $p_\mathrm{A,g}=\frac{1-\eta}{2}$, $M_\mathrm{A}=-\eta$. Thus, S$_\text{A}$ fully adapts the spin polarization of the magnetic tip: Depending on the voltage polarity, it is either driven into the ground or excited state. \\ 
\\
\textbf{Remote Spin}\\
For S$_\text{B}$ we consider the relaxation and excitation from the tip-sample processes with S$_\text{A}$ which have a finite probabilities to scatter S$_\text{B}$ caused by the small state mixing. We describe the rates by multiplying the transition matrix element of the combined system with the respective occupation of S$_\text{A}$ described in the section above. In addition, we include the relaxation process from sample-sample scattering as the next most relevant contribution: 
\\
\begin{equation}
\Gamma_\mathrm{exc} =\frac{G_{st}}{e} |V_\mathrm{DC}| \Big(p_\mathrm{A,g}P_{\mathrm{A},01} + p_\mathrm{A,e}P_{\mathrm{A},21}+p_\mathrm{A,e}P_{\mathrm{A},23}\Big)
\label{eq:sup_GammaExcB}
\end{equation}
\begin{equation}
\begin{aligned}
\Gamma_\mathrm{relax}
&= \frac{G_{st}}{e}\, |V_\mathrm{DC}|
\Big(
p_\mathrm{A,g} P_{\mathrm{A},10}
+ p_\mathrm{A,g} P_{\mathrm{A},12}
+ p_\mathrm{A,e} P_{\mathrm{A},32}
\Big) \\
&\quad + \frac{G_{ss}}{e^2}\, E_{10}
\Big(
p_\mathrm{A,g} P_{\mathrm{B},10}^{s\rightarrow s}
+ p_\mathrm{A,e} P_{\mathrm{B},32}^{s\rightarrow s}
\Big)
\end{aligned}
\label{eq:sup_GammaRelaxB}
\end{equation}\\
\noindent Here, the matrix elements $P_{\mathrm{A},if}$ are given by $P_{\mathrm{A},if}^{t \rightarrow s}$ ($P_{\mathrm{A},if}^{s \rightarrow t}$) for $V_\mathrm{DC}>0$ ($V_\mathrm{DC}<0$).
In the descriptions of Eq. \ref{eq:sup_GammaExcB} and \ref{eq:sup_GammaRelaxB} we identify two processes which when tunneling through the sensing spin excite or relax the remote spin (Fig. \ref{fig:sup_RSandFF}). One is a flip-flop interaction (FF) between S$_\text{A}$ and S$_\text{B}$ quantified by $P_{\mathrm{A},21}$ and $P_{\mathrm{A},12}$. The other is a direct scattering of S$_\text{B}$ quantified by $P_{\mathrm{A},01}$ ($P_{\mathrm{A},10}$) and $P_{\mathrm{A},23}$ ($P_{\mathrm{A},32}$) referred to as remote scattering (RS).\\
For the eigenstates found in section \ref{sec:sup_spinhamil} we can compute the relevant transition probabilities based on the matrix elements:\\
\begin{equation}
    P_{\mathrm{A},01}^{s\rightarrow t}=P_{\mathrm{A},23}^{s \rightarrow t} =
    P_{\mathrm{A},10}^{t\rightarrow s}=P_{\mathrm{A},32}^{t \rightarrow s} =
    \frac{1}{2}\frac{1+\eta}{2}\frac{1}{1+\alpha^2}
\end{equation}

\begin{equation}
    P_{\mathrm{A},10}^{s\rightarrow t}=P_{\mathrm{A},32}^{s \rightarrow t} =
    P_{\mathrm{A},01}^{t\rightarrow s}=P_{\mathrm{A},23}^{t \rightarrow s} =
    \frac{1}{2}\frac{1-\eta}{2}\frac{1}{1+\alpha^2}
\end{equation}

\begin{equation}
    P_{\mathrm{A},21}^{s\rightarrow t}=P_{\mathrm{A},12}^{s\rightarrow t} =
    P_{\mathrm{A},21}^{t\rightarrow s}=P_{\mathrm{A},12}^{t\rightarrow s} =
    \frac{1}{2}(\frac{\alpha}{1+\alpha^2})^2
\end{equation}

\begin{equation}
    P_{\mathrm{B},10}^{s\rightarrow s}=P_{\mathrm{B},32}^{s\rightarrow s}=
    \frac{1}{4}
\end{equation}\\ 
\noindent Here, we identify the transition matrix element for the flip-flop process $\mathcal{M}_\mathrm{FF}=(\frac{\alpha}{1+\alpha^2})^2=\frac{J_\perp^2}{4(\delta^2+J_\perp^2)}$ and the remote scattering process $\mathcal{M}_\mathrm{RS}=\frac{1}{1+\alpha^2}=\frac{1}{2}-\frac{\delta}{2\sqrt{\delta^2+J_\perp^2}}$ presented in Eq. \ref{eq:RateFF} and Eq. \ref{eq:RateRS}, respectively. They depend on the detuning $\delta$ and exchange coupling $J_\perp$ of the two spins. We note that we simplified $E_{10}=E_{32}\approx g\mu_\mathrm{B}B$, since $J_\parallel<B_\mathrm{tip}<B$.\\ 

\noindent For the magnetization of S$_\text{B}$ we then get:\\
\begin{equation}
M_\mathrm{B}=\frac{\eta V_\mathrm{DC}G_{st}\mathcal{M} + g\mu_\mathrm{B}B~G_{ss} / 2e^2}{|V_\mathrm{DC}|G_{st}\mathcal{M}+g\mu_\mathrm{B}B~G_{ss}/ 2e^2}
\end{equation}\\
\noindent With $\mathcal{M}=\mathcal{M}_\mathrm{RS}+\mathcal{M}_\mathrm{FF}$. For $G_\mathrm{set}\approx G_{st}$, where $G_\mathrm{set}=I/V_\mathrm{DC}$ is the setpoint conductance in the STM experiment, we can express $M_\mathrm{B}$ as a function of the induced tunneling current $I=G_{st}V_\mathrm{DC}$ which is independent of $G_\mathrm{set}$ and arrive at the formula given in the main text (Eq. \ref{eq:M_B}):\\
\begin{equation}
    M_\mathrm{B}(I) = \eta \frac{I-I_0}{|I|+|\eta I_0|}~~\text{with}~~I_0=-\frac{G_{ss} g \mu_\mathrm{B} B}{2e\mathcal{M} \eta}
\end{equation}\\
\noindent With the characteristic tunneling current $I_0$ describing the current needed to pump the system to $M_\mathrm{B}=0$.\\ 

\noindent In the limit of vanishing tunneling current for $V_\mathrm{DC} \rightarrow 0$, the occupation of the spin states is governed primarily by their intrinsic energy splitting rather than by current-induced transitions and Eq. \ref{eq:M_B} is no longer valid. Measurements performed at small $V_\mathrm{DC}$ therefore provide an upper bound of the effective spin temperature. For the experimental data presented in this work, the spin temperature was determined to be $T \approx 0.2~\mathrm{K}$ from the numerical solution of the full rate equation presented in section \ref{sec:sup_rates}.

\newpage
\section{Polarization Dependence}
\label{sec:sup_polarization}
\newpage
\section{Setpoint Dependence}
\label{sec:sup_setpoint}
\newpage
\section{Lifetime Estimation}
\label{sec:sup_lifetime}

\begin{figure}[!htbp]
  \centering
  \includegraphics[width=90mm]{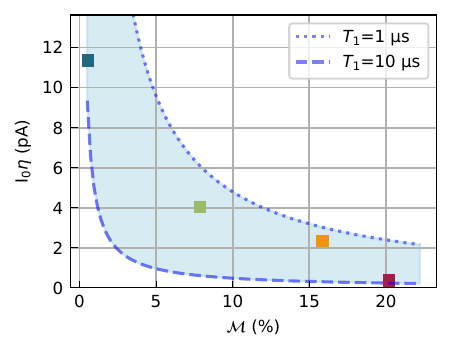}
  \caption{\textbf{Lifetime Estimation.} Plotting $I_0\eta$ as a function of $\mathcal{M}$. Squares are experimental values obtained from the measurements presented in Fig. \ref{fig:fig4} showing an inverse dependence. The dotted (dashed) line represents the behavior expected from the analytical approximation for $T_1=1~\mathrm{\mu s}$ ($T_1=10~\mathrm{\mu s}$).}
  \label{fig:sup_lifetime}
\end{figure}

\begin{figure}[!htbp]
  \centering
  \includegraphics[width=150mm]{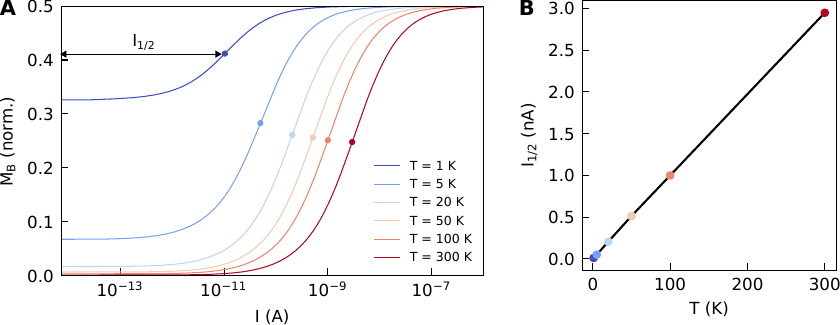}
  \caption{\textbf{Spin pumping for different temperatures.} (\textbf{A}) Magnetization of the remote spin $M_\mathrm{B}$ as a function of tunneling current $I$ for different temperatures. The simulation shows how the remote spin is pumped into a higher occupation of the ground state from thermal equilibrium. For higher temperatures more current is required to achieve $50~\%$ of the possible magnetization quantified by $I_{1/2}$ (dots along the line). (\textbf{B}) $I_{1/2} $ over $T$ extracted from the traces in (A) showing a linear behavior. The simulations demonstrate that with sufficient high currents ($\mathrm{nA}$ regime) spins can be initialized even at room temperature. [Simulation parameters: $g=2.0, B_\mathrm{tip}=36~\mathrm{mT},B=500~\mathrm{mT}, T=200~\mathrm{mK}, J_\parallel=-200~\mathrm{MHz}, J_\perp=300~\mathrm{MHz}, \eta=0.5, G_{st}=0.2~\mathrm{nS}, G_{ss}=5~\mathrm{nS}$].}
  \label{fig:sup_temperature}
\end{figure}

\begin{figure}[!htbp]
  \centering
  \includegraphics[width=100mm]{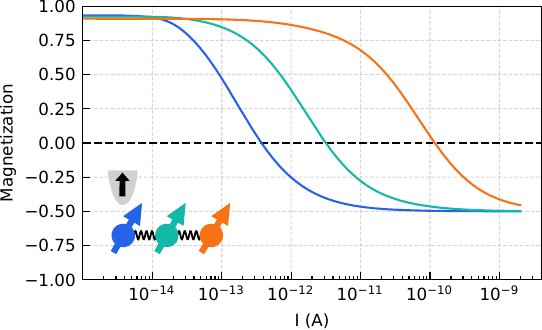}
  \caption{\textbf{Spin pumping simulation of a three spin chain.} Plotting the Magnetization of the three spins over the applied spin-polarized tunneling current $I$ (logarithmic scale). The geometry of the spin system is indicated by the sketch inserted at the lower left. Angular momentum is transferred from the tip which is acting on the first spin (blue) to the second (green) and third spin (orange). Detuning of the Zeeman energies is realized by different g-factors instead of a tip magnetic field. The results show that for a sufficient high tunneling current also a next nearest neighbour spin can be polarized. [Simulation parameters: $g_\mathrm{A}=2.0, g_\mathrm{B}=1.9, g_\mathrm{C}=1.8,B=500~\mathrm{mT}, T=200~\mathrm{mK}, J_\parallel=-200~\mathrm{MHz}, J_\perp=300~\mathrm{MHz}, \eta=-0.5, G_{st}=0.2~\mathrm{nS}, G_{ss}=5~\mathrm{nS}$].}
  \label{fig:sup_3spinchain}
\end{figure}

Since the pumping of the remote spin depends on the ratio between excitation via the spin-polarized tunneling current and relaxation via substrate electrons, one can estimate the lifetime of the remote spin in the absence of any pumping. This lifetime is given by:\\
\begin{equation}
    1/T_\mathrm{1} = \Gamma_\mathrm{B}^{s \rightarrow s} = 
    \frac{G_{ss}}{e^2}E_{10}\cdot P
\end{equation}\\
\noindent Where $P$ is the probability of an inelastic excitation. With the characteristic current $I_0=-\frac{G_{ss} g\mu_\mathrm{B}B}{2e\mathcal{M} \eta}$ we find:\\ 
\begin{equation}
    -I_0 \eta= \frac{e}{2P \cdot T_1}\frac{1}{\mathcal{M}}
\end{equation}\\
\noindent where we have used $E_\mathrm{10}=g\mu_\mathrm{B}B$. For the four investigated complex pairs presented in Fig. \ref{fig:fig4}, which exhibit different $\mathcal{M}$ values, we can plot this relation as shown in Fig. \ref{fig:sup_lifetime}. 
Using $P \approx17~\%$, which was determined for the Fe-FePc complex in previous work\cite{huang_quantum_2025}, we find that our measurements exhibit a spin lifetime in the range of $1\text{-}10~\mathrm{\mu s}$.\\ 

\clearpage
\section{Elevated Temperatures and Spin Chains}

\end{document}